\documentclass[twocolumn,letter]{jpsj3} 
\title{Magnetization Process 
of Kagome-Lattice Heisenberg Antiferromagnet}
\catcode`\@=11
\def\simle{\mathrel{\mathpalette\@versim<}}   
\def\simge{\mathrel{\mathpalette\@versim>}}   
\def\@versim#1#2{\lower2.5pt\vbox{\baselineskip0pt \lineskip-.5pt
   \ialign{$\m@th#1\hfil##\hfil$\crcr#2\crcr\sim\crcr}}}
\catcode`\@=12

\author{Hiroki \textsc{Nakano}
\thanks{E-mail address: hnakano@sci.u-hyogo.ac.jp} and 
Toru \textsc{Sakai}$^{1}$
\thanks{E-mail address: sakai@spring8.go.jp}}

\inst{Graduate School of Material Science, University of Hyogo,
Kouto 3-2-1, Kamigori, Ako-gun, Hyogo 678-1297, Japan \\
$^{1}$
Japan Atomic Energy Agency, SPring-8, 
Kouto 1-1-1, Sayo, Hyogo 679-5148, Japan
}

\recdate{\today}

\abst{The magnetization process 
of the isotropic Heisenberg antiferromagnet 
on the kagome lattice is studied.  
Data obtained from the numerical-diagonalization method 
are reexamined from the viewpoint of the derivative 
of the magnetization with respect to the magnetic field. 
We find that the behavior of the derivative 
at approximately one-third of the height 
of the magnetization saturation is markedly different 
from that for the cases of typical magnetization plateaux. 
The magnetization process of the kagome-lattice antiferromagnet 
reveals a new phenomenon, 
which we call the ``magnetization ramp.'' 
}

\kword{kagome lattice, 
antiferromagnetic Heisenberg spin model,
magnetization process, 
numerical-diagonalization method, Lanczos method}

\begin{document}
\maketitle

Frustration has attracted considerable attention 
as an origin of various exotic phenomena 
in condensed-matter physics. 
In particular, one of the typical frustrated systems 
in the field of magnetism is 
that due to the lattice structure based on the existence 
of neighboring bonds forming local triangles and tetragons. 
In such lattice structures, 
there is the case of the kagome lattice\cite{brief_history}.  
The $S=1/2$ kagome-lattice antiferromagnet has been 
studied; 
although it is believed that no long-range order is realized 
even at zero temperature owing to the strong frustration 
and large quantum fluctuation, 
no conclusive evidence has been obtained so far 
in spite of extensive studies. 
Under these circumstances, 
the kagome-lattice antiferromagnet has recently become 
a hot topic again because of the discovery of some new 
materials\cite{herbertsmithite_jacs,herbertsmithite_jpsj,
vesignieite_let,volborthite_jpsj_let,volborthite_prl}.  

From the theoretical point of view, however, 
it is well known that calculations of 
two-dimensional kagome-lattice systems are difficult 
even using a computational method. 
As reliable numerical methods, 
the quantum Monte Carlo (QMC) method, 
the density matrix renormalization group (DMRG) method, 
and the numerical-diagonalization method are well known. 
Although the DMRG method can treat systems with large sizes, 
the dimensionality of the systems is limited 
to being less than two. 
QMC simulations can treat systems in higher dimensions, 
but the so-called negative sign problem 
prevents us from obtaining reliable results  
in the cases of frustrated systems.  
Nevertheless, double-peak behavior was clarified 
to appear in the specific heat as a result 
of the great effort devoted toward carrying out 
effective sampling 
in the QMC simulations\cite{Nakamura_Miyashita}. 
Only the numerical-diagonalization method does not suffer 
from the limitation of dimensionality nor 
the negative sign problem.  
The disadvantage of the numerical-diagonalization method 
is the limitation that available system sizes are small. 
Thus, it is important 
to obtain precise results for systems that are as large 
as possible in the available calculations 
by the numerical-diagonalization method 
and 
to examine the finite-size effect very carefully, 
which contributes greatly the deep understanding 
of frustrated systems. 
With this background, 
numerical-diagonalization studies of kagome-lattice 
systems\cite{Lecheminant,Waldtmann,Cepas,Sindzingre} 
have been carried out. 

The magnetization process of the kagome-lattice 
Heisenberg antiferromagnet was examined 
by the numerical-diagonalization 
method\cite{Hida, Cabra, Honecker0, Honecker1}. 
In ref.~\ref{Hida}, the full process 
for system sizes of up to $N=30$ and 
the higher-field part of the process for $N=33$ were reported. 
Hida pointed out that a magnetization plateau appears 
at one-third of the height of the saturation. 
In ref.~\ref{Honecker1}, the magnetization process 
of the $N=36$ system was given; 
the authors of ref.~\ref{Honecker1} claimed that 
the existence of the plateau is {\it established} 
from their result. 


In this letter, we reexamine the magnetization process 
of the kagome-lattice Heisenberg antiferromagnet 
from another viewpoint in analyzing finite-size calculations. 
The point is to observe the field-derivative 
of the magnetization, 
namely, the differential magnetic susceptibility 
in magnetic fields. 
The observation will help us capture 
the behavior of the magnetization process with high accuracy. 
The purpose of this letter is to clarify, from the observation, 
that the magnetization process exhibits behavior 
different from those reported so far, for example, 
the magnetization plateau and magnetization cusp. 
We will call the new behavior of the magnetization process 
a ``magnetization ramp.'' 
For a system with size $N$, 
we can obtain the lowest energy $E(N,M)$ in each subspace 
for a given value of $S_z^{\rm tot}$ denoted by $M$ 
by the numerical diagonalization 
of the Lanczos algorithm and/or the householder algorithm, 
where $S_z^{\rm tot}$ represents 
the $z$-component of the total spin.  
We can evaluate the derivative by the expression 
\begin{equation}
\chi^{-1} = \frac{E(N,M+1)-2E(N,M)+E(N,M+1)}{1/M_{\rm sat}}, 
\label{chi_calc}
\end{equation}
where $M_{\rm sat}$ denotes the saturation of the magnetization, 
namely, $M_{\rm sat}=N S$ for the spin-$S$ system. 
Note that $\chi$ cannot be defined when 
one of the three $E$ in eq.~(\ref{chi_calc}) 
does not become the ground-state energy of the system 
in any magnetic field. 
We obtain $\chi$ in eq.~(\ref{chi_calc}) 
as a function of $M/M_{\rm sat}$. 


Before investigating the kagome-lattice antiferromagnet, 
let us observe numerical-diagonalization finite-size data 
of $\chi$ in one- and two-dimensional systems 
of interacting $S=1$ dimers.  
This model reveals a typical magnetization plateau 
at half the height of the saturation.   
\begin{figure}[tb]
\begin{center}
\includegraphics[width=4.5cm]{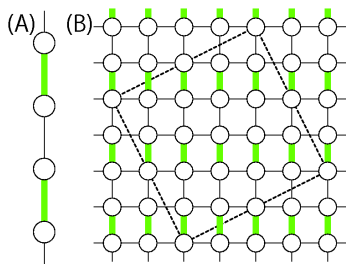}
\end{center}
\caption{Cluster shapes 
of the (A) one- and (B) two-dimensional systems 
of interacting $S=1$ dimers. 
Our calculations have been carried out for $N=20$ 
in both cases. 
In (B), the tilted square of $\sqrt{20}\times\sqrt{20}$ 
is shown with dotted lines. 
The thick green and thin black bonds denote 
the intra- and interdimer interactions, respectively. 
}
\label{fig1}
\end{figure}
\begin{figure}[tb]
\begin{center}
\includegraphics[width=7.0cm]{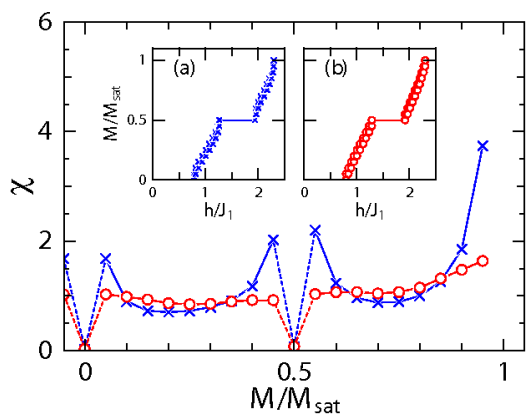}
\end{center}
\caption{Field-derivative of the magnetization $\chi$ 
of the one- and two-dimensional systems 
of interacting $S=1$ dimers, denoted 
by crosses and circles, respectively.  
Insets (a) and (b) show 
the magnetization process of the one-dimensional case (A) 
in Fig.~\ref{fig1} for $J_{2}/J_{1}=0.15$ and 
that of the two-dimensional case (B) for $J_{2}/J_{1}=0.05$, 
respectively. 
The main panel shows $\chi$ 
as a function of the magnetization divided by the saturation 
with corresponding colors and symbols. 
}
\label{fig2}
\end{figure}
The Hamiltonian of the interacting $S=1$ dimers is given by 
\begin{equation}
{\cal H} = \sum_{\langle i,j\rangle} J_{1}
\mbox{\boldmath $S$}_{i}\cdot\mbox{\boldmath $S$}_{j}
+\sum_{\langle i,j\rangle} J_{2}
\mbox{\boldmath $S$}_{i}\cdot\mbox{\boldmath $S$}_{j}
+ h \sum_{i} S^{z}_{i} ,
\end{equation}
where 
$\mbox{\boldmath $S$}_{i}$ denotes the $S=1$ spin operator. 
Here, the first term describes the intradimer interactions 
(denoted by the thick green bonds in Fig.~\ref{fig1}), 
the second term describes the interdimer interactions 
(denoted by the thin black bonds in Fig.~\ref{fig1}), 
and the third term is the Zeeman term. 
Note that $J_{1}$ is the energy unit; thus, we set $J_{1}=1$.  
We examine the cases 
for the clusters depicted in Fig.~\ref{fig1}. 
The results, 
which are depicted in Fig.~\ref{fig2}, 
will help us accurately capture 
the behavior of $\chi$ near the magnetization plateau. 
The system size is $N=20$ commonly. 
In the two-dimensional case, 
the tilted square shown in Fig.~\ref{fig1} is treated. 
The magnetization processes 
of the one-dimensional case (A) for $J_{2}/J_{1}=0.15$ and 
the two-dimensional case (B) for $J_{2}/J_{1}=0.05$ 
are presented in insets (a) and (b) in Fig.~\ref{fig2}, 
respectively. 
We choose these parameters so that 
the sum of the amplitudes of interdimer interactions 
operating a dimer of the thick green bond is 
the same in Figs.~\ref{fig1}(A) and \ref{fig1}(B). 
We can clearly observe a plateau at $M/M_{\rm sat}=1/2$ 
in both cases. 
Note that the experimentally observed magnetization processes 
were reported for one-dimensional systems 
of Ni compounds\cite{Narumi1,Narumi2} and 
for the two-dimensional system of the organic biradical 
F$_{2}$PNNNO\cite{Hosokoshi}. 
The one- and two-dimensional magnetization processes 
appear to be very similar; 
it is difficult to find a clear difference between them. 
Let us next discuss the behavior of $\chi$,   
the results for which are presented in the main panel. 
Near half the height of the saturation, 
we can observe that 
$\chi$ diverges in the one-dimensional system, 
while it remains finite in the two-dimensional system. 
These properties originate from the nature of the density 
of states determined from the parabolic dispersion. 
It is noticeable that 
the behavior does not change for each dimension, 
irrespective of the difference 
in the lower- or higher-field sides, 
and that the value of $\chi$ just at $M$ of the plateau 
is discontinuous with the values of $\chi$ around it. 
One can find that $\chi$ clearly exhibits the behavior 
of a typical magnetization plateau from the above observation. 
Note that these characteristics of $\chi$ appear not only 
at $M/M_{\rm sat}=1/2$ but also at $M=0$. 
The behavior at $M=0$ is already well known in various 
cases\cite{MTakahashi_TSakai,Affleck,TSakai_MTakahashi,
Katoh_Imada}; 
the behavior of $\chi$ in these cases clearly 
satisfies the characteristics mentioned above. 


\begin{figure}[tb]
\begin{center}
\includegraphics[width=6.0cm]{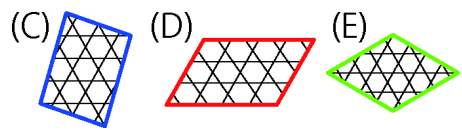}
\end{center}
\caption{
Shapes of the finite-size clusters 
in the kagome lattice. 
Cluster (C) is the same as that for $N=33$ 
in ref.~\ref{Hida}. 
Cluster (D) is the new cluster for $N=36$, 
while cluster (E) is that for $N=36$ 
in ref.~\ref{Honecker1}. 
}
\label{fig3}
\end{figure}
\begin{figure}[h]
\begin{center}
\includegraphics[width=8.5cm]{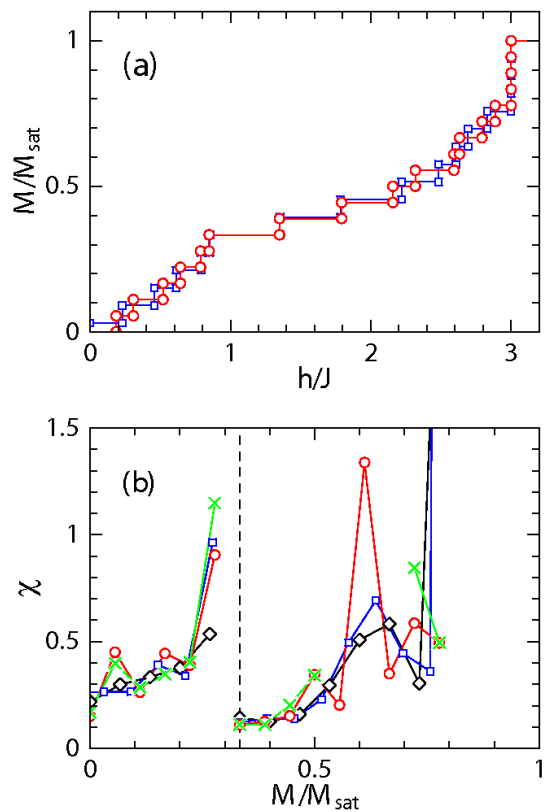}
\end{center}
\caption{Results for the kagome-lattice antiferromagnets. 
(a) Magnetization processes shown
for the $N=36$ cluster (D) and 
the $N=33$ cluster (C) with circles and squares, respectively. 
(b) Field-derivative of the magnetization $\chi$ 
as a function of the magnetization divided by the saturation. 
Circles and squares in (b) denote the results of $\chi$ 
corresponding to the cases in (a). 
Crosses correspond to the $N=36$ cluster (E). 
Note that part of $\chi$ for the $N=36$ cluster (E) is missing 
because $\chi$ cannot be defined when the lowest-energy state 
in the subspace does not become the ground state of the system 
in the magnetic field. 
Diamonds represent the case of the $N=30$ cluster 
investigated in ref.~\ref{Hida}. 
}
\label{fig4}
\end{figure}
We now examine the magnetization process 
of the kagome-lattice antiferromagnet. 
Its Hamiltonian is given by 
\begin{equation}
{\cal H} = \sum_{\langle i,j\rangle} J
\mbox{\boldmath $S$}_{i}\cdot\mbox{\boldmath $S$}_{j}
+ h \sum_{i} S^{z}_{i} ,
\label{H_kagome}
\end{equation}
where the sum of the first term runs over 
the nearest neighbors on the kagome lattice. 
Note that $\mbox{\boldmath $S$}_{i}$ in eq.~(\ref{H_kagome}) 
denotes the $S=1/2$ spin operator 
and that $J$ is the energy unit; thus, we set $J=1$.  
The shape of the finite-size cluster of the kagome lattice 
under the periodic boundary condition 
is not necessarily unique 
even when the number of spin sites $N$ is given. 
In fact, for $N=36$, there is another shape, 
denoted by the red cluster (D) in Fig.~\ref{fig3}, 
which is different from that treated by Honecker {\it et al}. 
in ref.~\ref{Honecker1}; 
the cluster of Honecker {\it et al}. 
is shown in Fig.~\ref{fig3}(E).  
In this letter, we present the result for the red cluster (D) 
for $N=36$, which helps us know the finite-size effects 
from the difference between the two cases (D) and (E) for $N=36$. 

The result of the magnetization process 
of the red cluster (D) for $N=36$ sites is depicted 
in Fig.~\ref{fig4}(a). 
We also give the full result of the blue cluster (C) for $N=33$, 
which is the same cluster as that treated by Hida 
in ref.~\ref{Hida}. 
In Fig.~\ref{fig4}(b), we present 
the results of $\chi$ for the finite-size systems 
of $N=30$, 33, and 36. 
We focus our attention on 
the behavior of $\chi$ around $M/M_{\rm sat}\sim 1/3$. 
One can observe that 
around $M/M_{\rm sat}\sim 1/3$, shown in Fig.~\ref{fig4}(b), 
$\chi$ exhibits divergent behavior on the smaller-$M$ side, 
while it is very small on the larger-$M$ side; 
namely, the behavior on the smaller-$M$ side 
is not the same as that on the larger-$M$ side. 
This fact is markedly different from the behavior of $\chi$ 
for the interacting $S=1$ dimer systems 
around $M/M_{\rm sat}\sim 1/2$ in Fig.~\ref{fig2}. 
In particular, 
the divergent behavior on the smaller-$M$ side 
appears to be similar to that in the one-dimensional case 
even though the lattice is two-dimensional without any isotropy.  
This behavior may suggest that a one-dimensional spin structure 
forms in the two-dimensional lattice structure. 
A similar one-dimensional feature 
in the kagome lattice was reported 
in refs.~\ref{Ohashi1} and \ref{Ohashi2}. 
The relationship between the present result and 
that in refs.~\ref{Ohashi1} and \ref{Ohashi2} is an open 
problem which should be tackled in the near future. 
Since the divergent behavior appears on the smaller-$M$ side,  
the external field corresponding to $M/M_{\rm sat}\sim 1/3$ 
is the critical field.  
On the other hand, on the larger-$M$ side, 
there is no divergent behavior of $\chi$. 
It is noticeable that $\chi$ 
from the larger-$M$ side is continuous 
with $\chi$ just at $M/M_{\rm sat}=1/3$. 
It is unclear whether or not $\chi$ at $M/M_{\rm sat}=1/3$ 
vanishes as $N\rightarrow\infty$ at the present time. 

\begin{figure}[tb]
\begin{center}
\includegraphics[width=6.3cm]{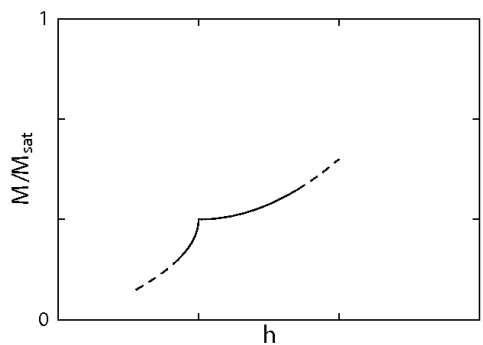}
\end{center}
\caption{Schematic shape of the magnetization ramp.}
\label{fig5}
\end{figure}

From these observations, the behavior of the magnetization 
around $M/M_{\rm sat}=1/3$ is anomalous; 
it is reasonable to consider that 
the behavior is a new phenomenon in magnetization. 
The schematic behavior of $M$ around $M/M_{\rm sat}=1/3$ 
based on the discussion of the characteristics of $\chi$ 
is shown in Fig.~\ref{fig5}. 
The shape is similar to the ramp of a Nordic ski jump 
when seen from the horizontally sideways direction. 
Thus, we call the behavior of $M$ a magnetization ramp. 
This behavior is markedly different 
from the magnetization process of a classical system 
on the kagome lattice\cite{Cabra}. 
Although the magnetization ramp may appear 
to be a magnetization cusp, the ramp is definitely 
different from a cusp 
because a conventional cusp occurs at the level crossing 
between two states with finite gradients, 
for example, as reported in ref.~\ref{Okunishi}. 
Thus, a magnetization ramp is 
an unusual and new phenomenon. 

Next, we discuss the characteristic behavior of other parts 
with respect to the relative height 
in our finite-size magnetization processes. 
We successfully observe a jump near the saturation, 
which was proven in ref.~\ref{Zhito_Tsunetsugu}. 
Note that several eigenstates with various values 
of $S_z^{\rm tot}$ degenerate just at $h/J=3$. 
In the $N=36$ magnetization process reported 
in ref.~\ref{Honecker1}, a jump appears at $M/M_{\rm sat}=11/18$; 
at the field where the lowest-energy eigenvalue 
in the subspace of $S_z^{\rm tot}=10$ meets 
that of $S_z^{\rm tot}=12$, 
the energy is lower than the lowest-energy eigenvalue 
in the subspace of $S_z^{\rm tot}=11$. 
In this meaning, this jump is different from the jump 
near the saturation discussed above. 
In the present result for $N=36$ in Fig.~\ref{fig4}(a), 
however, the jump at $M/M_{\rm sat}=11/18$ does not appear.  
It is thus unclear 
whether or not the jump at this relative height 
survives in the thermodynamic limit. 
Even in these situations, 
when one focuses on the behavior of $\chi$ 
near $M/M_{\rm sat}\sim 0.6$ in Fig.~\ref{fig4}(b), 
$\chi$ is enhanced irrespective of the system size 
except for the case of the green cluster (E) for $N=36$, 
in which $\chi$ cannot be obtained near $M/M_{\rm sat}\sim 0.6$; 
the enhancement suggests the existence of some anomaly 
around $M/M_{\rm sat}\sim 0.6$. 
It should be clarified in future studies 
whether this anomaly is a jump 
in the magnetization process or another. 


Finally, let us discuss the relationship 
between our observation of the magnetization process 
of the ideal $S=1/2$ kagome-lattice Heisenberg antiferromagnet  
and the magnetization measurement 
of the actual compounds volborthite and vesignieite, 
which have relatively small values of $J$. 
These materials include other factors 
that are not considered 
in the ideal situation assumed in this study. 
Vesignieite includes a small amount of impurity spins 
according to the results of recent studies. 
In volborthite, on the other hand, there is 
spatial anisotropy in the bonds, 
although very pure samples are available. 
It should be noted that 
the magnetization processes of both materials 
reveal flat behavior 
at about $M/M_{\rm sat}\sim 0.4$\cite{private_comm_okamoto}, 
which is slightly larger 
than the (theoretical) numerical result of $M/M_{\rm sat}=1/3$. 
It should be investigated in the near future 
whether or not 
the missing factors mentioned above are the origin of this difference in height of $M$ exhibiting the flat behavior and 
whether or not these factors affect 
the appearance of the magnetization ramp. 
In particular, in ref.~\ref{volborthite_jpsj_let} 
it was reported that 
in volborthite, another anomalous behavior 
in its magnetization process is observed 
at around $M=(1/6) \mu_{\rm B}$/Cu and (1/45)$\mu_{\rm B}$/Cu, 
referred to 
as ``magnetization steps.''\cite{note_magnetization_step}
In the result for the red cluster (D) of $N=36$ 
in Fig.~\ref{fig4}, 
small enhancements of $\chi$ appear at $M/M_{\rm sat}=1/6$ 
and 1/18. 
Unfortunately, the resolution is not sufficient 
to judge whether or not the finite-size data surely 
capture the behavior of the magnetization steps; 
calculations for larger systems are greatly anticipated. 


In summary, we have examined the magnetization process 
of the $S=1/2$ kagome-lattice Heisenberg antiferromagnet 
from the viewpoint of its gradient $\chi$ 
based on numerical-diagonalization finite-size data.  
The observation of $\chi$ is very useful 
for extracting characteristics in the magnetization 
process that are difficult to detect 
when only the magnetization process is observed. 
Our observation leads to a new phenomenon, 
a magnetization ramp. 
To confirm the existence of the phenomenon, 
two approaches will be necessary in future studies. 
One of them is the construction of an effective theory. 
The other is a numerical examination of systems 
with larger sizes. 
We plan to carry out calculations for $N=39$; 
the results will be published elsewhere. 
In such works, something unknown 
in properties of kagome-lattice systems 
may be clarified further. 

\section*{Acknowledgments}
We wish to thank 
Professor K.~Hida, 
Professor T.~Tonegawa, 
Professor S.~Suga, 
Professor M.~Isoda, 
Professor Z.~Hiroi, 
Professor M.~Tokunaga, 
and  
Dr.~Y.~Okamoto
for fruitful discussions. 
This work was partly supported by Grants-in-Aid (No.~20340096), 
Grants-in-Aid for Scientific Research and Priority Areas 
``Physics of New Quantum Phases in Superclean Materials,'' 
``High Field Spin Science in 100T,'' and 
``Novel States of Matter Induced by Frustration'' 
from the Ministry of Education, Culture, Sports, Science 
and Technology of Japan. 
Part of the computations were performed using the facilities 
of the Information Initiative Center, Hokkaido University; 
the Information Technology Center, Nagoya University; 
Department of Simulation Science, 
National Institute for Fusion Science; 
and the Supercomputer Center, 
Institute for Solid State Physics, 
University of Tokyo. 
Computations of diagonalization 
without process parallel calculations 
were based on TITPACK ver.2 coded by H. Nishimori. 



\begin{thebibliography}{99}
\bibitem{note_magnetization_step}
The relative heights $M/M_{\rm sat}$ 
of the magnetization steps in ref. \ref{volborthite_jpsj_let} 
are the consequence of the assumption 
that the $g$-factor is 2. 
Note that the actual $g$-factor is 
slightly different from $g=2$ \cite{Ohta}.
\end{thebibliography}
\end{document}